\documentclass[twocolumn,showpacs,pra,groupedaddress,superscriptaddress,nofootinbib,longbibliography]{revtex4-1}

\usepackage{graphicx,amsmath,amsthm,amssymb,amsbsy,subfigure,hyperref,bbm,times,txfonts,float}
\usepackage{subfigure,hyperref,bbm,times}
\usepackage[T1]{fontenc}
\usepackage{braket}
\usepackage{epsfig}
\usepackage{color}
\usepackage{graphicx}
\usepackage{dcolumn}
\usepackage{bm}

\providecommand{\openone}{\leavevmode\hbox{\small1\kern-3.8pt\normalsize1}}

\usepackage{soul}

\hypersetup{
	colorlinks=true,
	linkcolor=blue,
}
\graphicspath{{figure/}}

\begin{document}
	
	
	\title{Searching for exceptional points and inspecting non-contractivity of trace distance  in (anti-)$\mathcal{PT}\!-$symmetric systems}
	
	\author{Hossein Rangani Jahromi}
	\email{h.ranganijahromi@jahromu.ac.ir}
	\affiliation{Physics Department, Faculty of Sciences, Jahrom University, P.B. 74135111, Jahrom, Iran}

	\author{Rosario Lo Franco}%
	\email{rosario.lofranco@unipa.it}
	\affiliation{Dipartimento di Ingegneria, Universit\`{a} di Palermo, Viale delle Scienze, Edificio 6, 90128 Palermo, Italy}%
	
	\date{\today}
\begin{abstract}
	Non-Hermitian systems with parity-time ($\mathcal{PT}$) symmetry and anti-$\mathcal{PT}$ symmetry give rise to exceptional points (EPs) with intriguing properties related to, e.g., chiral transport and enhanced sensitivity, due to the coalescence of eigenvectors.  
	In this paper,  we propose a powerful and easily computable tool, based on the Hilbert-Schmidt speed (HSS), which does not require the diagonalization of the evolved density matrix, to  detect exactly the EPs and hence the critical behavior of the (anti-)$\mathcal{PT}\!-$symmetric systems, especially high-dimensional ones. Our theoretical predictions, made without the need for modification of the Hilbert space, which is performed by diagonalizing one of the observables,  are completely consistent with results extracted from recent experiments studying the criticality in (anti-)$\mathcal{PT}\!-$symmetric systems. Nevertheless, not modifying the Hilbert space of the  non-Hermitian system, we find that the trace distance, a measure of distinguishability of two arbitrary quantum states, whose dynamics is known as
	a faithful witness of non-Markovianity in Hermitian systems, may be
	non-contractive under the non-Hermitian evolution of the system. Therefore, it lacks one of the most  important characteristics which must be met by any standard witness of non-Markovianity. 
\end{abstract}


\maketitle


\section{Introduction \label{introduction}}
One of the fundamental postulates of   quantum theory is that the Hamiltonian of an isolated system is Hermitian. 
This Hermiticity seems to be a compelling postulate because it ensures that the eigenvalues of the Hamiltonian are real. Moreover, a Hermitian Hamiltonian leads to a unitary time evolution and consequently the conservation of probability. However,  decoherence effects  are ubiquitous in nature because the physical systems are always inevitably affected by the environment. Under this condition, the dynamics of the system is dominated by a Lindbladian generalized master equation. Because accurately solving this master equation is usually a very difficult task, especially in high-dimensional systems, an approximate but fruitful approach \cite{nielsen2010quantum} to describe the dynamics of open quantum systems is  directly handling the Schr\"{o}dinger equation $ i\partial_{t} \ket{\psi}=H \ket{\psi}$,
such that the time evolution operator is determined by an effective time-dependent Hamiltonian $ H $ which is not necessarily Hermitian \cite{daley2014quantum,dalibard1992wave}. Non-Hermitian Hamiltonians usually have complex  eigenvalue spectra and do not conserve probabilities, and therefore they often only serve as phenomenological descriptions of an open quantum system. Nevertheless,  there is a certain class of non-Hermitian Hamiltonians, invariant under a combination of parity ($ \mathcal{P} $) and time-reversal ($ \mathcal{T}  $) operations, i.e., $ [\mathcal{PT},H] =0$, whose spectrum can be entirely real as long as they respect  $ \mathcal{PT}  $ symmetry
\cite{bender2002complex,bender2003must,ahmed2001real,bender2007faster,znojil2001pt}.
Some applications and features of
$ \mathcal{PT}  $ symmetry are addressed in Refs. \cite{el2018non,feng2017non,jing2015optomechanically,zhang2015giant,quijandria2018pt,arkhipov2019scully,huybrechts2020validity,zhang2020breaking,arkhipov2020liouvillian,PhysRevA.98.022117,PhysRevA.101.033820}.

In general, $ \mathcal{PT} \!-$symmetric systems exhibit two phases: the unbroken phase in which the entire eigenspectrum is real, and the
broken phase where some or all of the eigenvalues form complex
conjugate pairs. This \textit{phase transition}
occurs at an special point where $ n $ eigenvalues, as
well as their corresponding eigenvectors, coalesce \cite{kato2013perturbation,heiss2012physics,ozdemir2019parity}.
This non-Hermitian degeneracy, also known as exceptional point (EP) of order n (EPn),
has recently emerged as a new way to engineer the response of the open
physical system. 

In optics, the abrupt nature
of the phase transitions being encountered around EPs has been demonstrated to lead to many intriguing phenomena, such
as unidirectional invisibility \cite{lin2011unidirectional,peng2014parity}, loss-induced transparency \cite{guo2009observation}, band
merging \cite{zhen2015spawning,makris2008beam}, laser mode selectivity \cite{hodaei2014parity,feng2014single}, topological chirality \cite{doppler2016dynamically,xu2016topological}, new types of thresholdless phonon lasers \cite{jing2014pt,lu2017exceptional}, and even exceptional photon
blockade \cite{huang2020exceptional}.
These important phenomena have been experimentally observed in different platforms based on optomechanics \cite{jing2014pt}, electronics \cite{schindler2011experimental}, metamaterials \cite{kang2013effective}, acoustics \cite{zhu2014p,fleury2015invisible},
and plasmonics \cite{benisty2011implementation}.

Moreover,  recently, it has been demonstrated that the bifurcation properties of
second-order non-Hermitian degeneracies can be used as an efficient tool to improve
the sensitivity (frequency shifts) of resonant optical
structures to external perturbations \cite{wiersig2014enhancing}. In addition, it is of particular interest \cite{hodaei2017enhanced} to use even higher-order EPs (greater than second order), which in principle could considerably amplify the effect of perturbations, leading to  greater sensitivity beyond what is possible in standard
arrangements \cite{wiersig2020review}.

Recently,  anti-$ \mathcal{PT} \!-$symmetric systems where the Hamiltonian is anti-commutative with
the joint $\mathcal{PT}$ operator, i.e.,  $ \{H,\mathcal{PT}\}=0 $, have attracted much research
interest. Interesting physical phenomena reported in anti-$\mathcal{PT}\!-$symmetric systems include optical
systems with constant refraction \cite{yang2017anti} and optical
materials with balanced positive and negative index \cite{ge2013antisymmetric}. Moreover,  several relevant experiments have been realized in  diffusive systems \cite{li2019anti}, 
electrical circuit resonators, and atomic \cite{peng2016anti,chuang2018realization,wang2016optical} or molecular systems \cite{jingwei2020observation}. Additionally, quantum processes such as the observation of EPs \cite{choi2018observation}, symmetry-breaking transition \cite{peng2016anti}, and simulation of anti-$\mathcal{PT}\!-$symmetric Lorentz dynamics \cite{li2019experimental}, which are important phenomena different from Hermitian quantum mechanics, have been addressed in these experiments.

Decoherence control is a key task for practical implementations of nanoscale solid-state quantum information processing, in which the decoherence may be mainly affected by non-Markovian dynamics due to the strong backactions from the environment \cite{breuer2002theory,lidar2019lecture,chen2015using,chen2009detecting,yin2012spin,zhang2012general,
	chen2016quantifying,lu2020observing,jahromi2019quantum,jahromi2019multiparameter,PhysRevA.95.052126,jahromi2015precision,NMexp5,LoFrancoNatCom,darrigo2012AOP,
	GholipourAnnPhys,mortezapourOpt,lofrancoreview,bellomo2007non,Mortezapour_2017,NM1,NMexp1,
	NMexp4,PhysRevA.85.032318,PhysRevA.92.012315,naikoo2019facets,PhysRevA.92.032311,teittinen2018revealing,D_Arrigo_2013,Bellomo_2012}. A fundamental issue is how to accurately define Markovianity in quantum processes \cite{breuer2016colloquium,rivas2014quantum}. In order to present a definition of a Markov process in the quantum regime, it is essential to understand the concept of
Markov process in the classical scenario.

\par
A classical \textit{Markov process} 
is defined as a family of random
variables $ \{X(t), t \in I \subset \mathbb{R}\} $, for which the probability that $  X $ takes a value $ x_{n} $ at any arbitrary time $ t_{n} \in I$,
if it took  value $ x_{n-1} $ at some previous time
$ t_{n-1} < t_{n} $, can be uniquely determined, and  is not influenced
by the possible values of $ X $ at previous times to $ t_{n-1} $. One can 
formulate it in terms of conditional probabilities as follows: $	\mathbb{P}(x_{n}, t_{n}|x_{n-1}, t_{n-1};...; x_{0}, t_{0}) = \mathbb{P}(x_{n}, t_{n}|x_{n-1}, t_{n-1})$ for all $\{t_{n} \geq  t_{n-1} \geq  ... \geq  t_{0}\} \subset I$.
Roughly speaking,
its concept is informally encapsulated by the statement that "a
Markov process has no memory of the history of past
values of $ X $". 

\par
In order to obtain a similar formulation  
in the quantum realm we demand a way to define
$ \mathbb{P}(x_{n}, t_{n}|x_{n-1}, t_{n-1};...; x_{0}, t_{0}) $ for quantum systems. In the classical scenario we can sample a  random
variable without affecting its posterior statistics. However, 'sampling'
a quantum system requires measuring process,  disturbing the state
of the system and affecting the subsequent outcomes. Therefore,
$ \mathbb{P}(x_{n}, t_{n}|x_{n-1}, t_{n-1};...; x_{0}, t_{0}) $  depend on both the
dynamics and the measurement process. Because in such a case the
Markovian character of a quantum dynamical system would
depend on the the measurement scheme  chosen to achieve
$ \mathbb{P}(x_{n}, t_{n}|x_{n-1}, t_{n-1};...; x_{0}, t_{0}) $, a definition
of quantum Markovianity in terms of which is not an easy task.
In fact,
the definition of Markovianity should be independent of what
is required to verify it. 

\par
The  aforementioned problem can be solved by adopting a  different approach
focusing on the study of one-time probabilities $ \mathbb{P}(x,t) $. In \textit{linear} quantum evolutions it may  lead to concept of \textit{divisibility} which can be defined  without any explicit reference to measurement processes in the quantum
realm \cite{rivas2014quantum}. \textit{Although these probabilities help us to  
	avoid the difficulties
	associated with the measurement disturbance, their efficiency in the scenarios involving some  measurement with postselection is controversial.}

One of the most well-known approaches to identify the non-Markovian character of the system dynamics has been proposed by Breuer-Laine-Piilo (BLP), namely the distinguishability of two evolving  states of the  quantum system \cite{breuer2009measure,laine2010measure}. For two arbitrary states  $ \rho_{1} $ and $ \rho_{2} $, this distinguishability is quantified by the trace distance (TD) $D(\rho_{1},\rho_{2})=\frac{1}{2}\text{Tr}|\rho_{1}-\rho_{2}|$, where $ |A|=\sqrt{A^{\dagger}A} $ for some operator $ A $.  

To explain the physical origin of this
interpretation we consider two parties, Alice and Bob, and
assume  that Alice prepares a quantum system in one of
two states $ \rho_{1} $ or $ \rho_{2} $, with a probability of $ 1/2 $ each. Then,  the system is passed into a "black box" where it may be
probed by
Bob  in any way allowed by the laws of quantum mechanics. Bob's task  is to determine whether the system is in the state
$ \rho_{1} $ or $ \rho_{2} $, by
means of a
\textit{single} quantum measurement.
It has been  shown that the maximal success probability which
Bob can achieve through an optimal strategy is directly
related to the trace distance \cite{fuchs1998information,breuer2016colloquium}: $ 	P_{\text{max}}=1/2\bigg(1+D(\rho_{1},\rho_{2})\bigg). $
Therefore,  the trace distance denotes the bias in favor
of a correct state discrimination by Bob, and hence it
can be interpreted as the distinguishability of the quantum states $ \rho_{1} $ and $ \rho_{2} $.
\par
The TD   is  contractive under completely positive and trace-preserving (CPTP) maps, i.e. $ D(\mathcal{E}_{t}(\rho_{1}),\mathcal{E}_{t}(\rho_{2}))\leq D(\rho_{1},\rho_{2})$, if  $\mathcal{E}_{t}$ is a CPTP map.
Therefore,  a monotonic decrease in the distinguishability ($ \dot{D} <0 $) indicates unidirectional information flow from the system to the environment. However,  an increase in the distinguishability ($ \dot{D} >0 $) signifies  backflow of information from the environment to the system, indicating that the time evolution of the system is affected by the history of system-environment interaction. This is one of the most well-known definitions of quantum  non-Markovianity in the literature.  It should be noted that there are also other ways to define and 
detect non-Markovianity or memory effects in quantum mechanics (see \cite{rivas2014quantum} for a review). It should be noted that non-Markovian dynamics is always associated with non-unitary evolution of the system. Contrary to Hermitian systems which can evolve unitarily and non-unitarily,  (anti-)$\mathcal{PT}\!-$ dynamics is intrinsically nonunitary both
in the unbroken and broken phases \cite{kawabata2017information}, satisfying one of the necessary conditions to exhibit non-Markovian dynamics.

\par Similar to classical scenario, appearance of a measurement  with postselection  in the process may call into question the validity of 
this non-Markovianity definition, because the time evolution of  the system becomes dependent on not only the  history of system-environment interaction but also the output of the postselection. \textit{This dependence of the quantum non-Markovianity definition on  the measurement
scheme, chosen to achieve the time evolution, is bothering even if we distinguish between memory effects and backflow of information from the environment.} Following this idea,  we show that the BLP measure does not have a necessary condition to  be a figure of merit for  characterizing   non-Markovianity, if it is definable in  dynamics involving postselection.
To this aim, we investigate the TD contractivity, a necessary condition which should be satisfied by this measure when it is used for  defining  non-Markovianity.

Recently, the Hilbert-Schmidt speed (HSS), a measure of quantum statistical speed not requiring diagonalization of the system reduced density matrix, has been introduced as a faithful witness of non-Markovianity in Hermitian systems, completely consistent with the BLP witness \cite{jahromi2020witnessing}. Moreover, the HSS has been introduced  as an efficient figure of merit 
for quantum estimation of phase 
encoded into the initial state 
of open $ n-$qubit systems \cite{jahromi2021hilbert}. Possibility to  enhance  quantum sensing near the EPs \cite{wiersig2014enhancing,chen2017exceptional,yu2020experimental,zhang2019quantum} and application of the TD to characterize the criticality in (anti-)$ \mathcal{PT} \!-$symmetric systems \cite{kawabata2017information,xiao2019observation,jingwei2020observation} motivate one to investigate the efficiency of the HSS measure to find  these singular points in non-Hermitian systems. Moreover, they motivate us to study the  relationship among the HSS, TD, and quantum Fisher information (QFI),  playing a central role in quantum estimation theory, in (anti-)$ \mathcal{PT} \!-$symmetric systems. 

\par In this paper, we address the aforementioned issue and also \emph{show how the HSS can effectively determine the EPs and reveal the critical behavior of non-Hermitian (anti-)$ \mathcal{PT}  \!-$symmetric systems}. Because finding the coalescence of eigenvectors at EPs by numerical
full diagonalization, required when computing the TD, may be  a tedious and time-consuming chore, this theoretical development for detecting the EPs could be very useful.

This paper is organized as follows. In Sec.~\ref{HILBERT-SCHMIDT SPEED} we briefly
review the definition of the Hilbert-Schmidt speed. In Sec.~\ref{evolution} the time evolution of 
the (anti-)$ \mathcal{PT}  \!-$symmetric systems as well as non-contractivity of the trace distance  are discussed. In Sec.~\ref{sec:HSSnonHermitianMeasure}
we propose a protocol based on  the HSS to characterize phase transitions in (anti-)$ \mathcal{PT}  \!-$symmetric systems. The efficiency of this protocol for detecting EPs and criticality in single-qubit systems is discussed in Secs.~\ref{SECTIONEXAMPLES} and \ref{ANTISECTIONEXAMPLES}. Moreover, the sensitivity of this witness is also studied for a high-dimensional $ \mathcal{PT}  \!-$symmetric system in Sec.~\ref{qudit}. Finally, Sec.~\ref{conclusion} summarizes the main results and prospects.

\section{HILBERT-SCHMIDT SPEED}\label{HILBERT-SCHMIDT SPEED}

The distance measure, defined as \cite{gessner2018statistical}   
\begin{equation}\label{cdis}
	[d(p,q)]^{2}=\dfrac{1}{2}\sum\limits_{x}^{}|p_{x}-q_{x}|^{2},
\end{equation}
where $ p = \{p_{x}\}_{x} $ and $ q = \{q_{x}\}_{x} $ are probability distributions, leads to the classical statistical speed
\begin{equation}\label{cspeed}
	s[p(\varphi_{0})]=\dfrac{\mathrm{d}}{\mathrm{d}\varphi}\ d(p(\varphi_{0}+\varphi),p(\varphi_{0})).
\end{equation}
Thus, one can define a special kind of quantum statistical speed called the HSS by extending these classical notions to the quantum case. 
To this aim, we may consider  a given pair of quantum states $ \rho $ and $ \sigma $, and  write $ p_{x} = \text{Tr}\{E_{x}\rho\} $ and $ q_{x} = \text{Tr}\{E_{x}\sigma\} $ denoting the measurement
probabilities corresponding to the positive-operator-valued measure (POVM) defined by the $ \{E_{x}\geq 0\} $ which satisfies $\sum\limits_{x}^{} E_{x} = \mathbb{I}  $.
Then the associated quantum distance called  the Hilbert-Schmidt  distance $ \delta_{HS} $ \cite{ozawa2000entanglement} can be achieved by maximizing the classical distance of Eq.~(\ref{cdis}) over all possible choices of POVMs \cite{PhysRevA.69.032106}
\begin{equation}\label{qdistance}
	\delta_{HS} (\rho,\sigma)\equiv \max_{\{E_{x}\}}\text{d}(\rho,\sigma)=\sqrt{\frac{1}{2}\text{Tr}[(\rho-\sigma)^{2}]}.
\end{equation}
Consequently, the HSS, the corresponding quantum statistical speed, is obtained by maximizing the classical statistical speed of Eq.~(\ref{cspeed}) over all possible POVMs \cite{paris2009quantum,gessner2018statistical}
\begin{align}\label{HSSS}
	HS\!S \big(\rho(\varphi)\big)\equiv HS\!S_{\varphi}
	&\equiv \text{S}\big[\rho(\varphi)\big]=\max_{\{E_{x}\}} \text{s}\big[p(\varphi)\big]\nonumber\\
	&=\sqrt{\frac{1}{2}\text{Tr}\bigg[\bigg(\dfrac{\mathrm{d}\rho(\varphi)}{\mathrm{d}\varphi}\bigg)^2\bigg]},
\end{align}
which can be easily computed without  diagonalizing  $ \text{d}\rho(\varphi)/\text{d}\varphi $.

 \section{Time evolution of the system governed by a (an)  (anti-)$ \mathcal{PT} \!-$symmetric Hamiltonian leading to non-contractivity of TD }\label{evolution}
 We directly apply the conventional quantum mechanics on (anti-)$ \mathcal{PT} \!-$symmetric systems to obtain the evolved state.
 Accordingly, the dynamics governed by a (an) (anti-)$ \mathcal{PT} \!-$symmetric system with non-Hermitian Hamiltonian $ H_{\mathcal{PT}}^{(anti)} $ is described by \cite{brody2012mixed,kawabata2017information}
 \begin{equation}\label{densityR}
 	\rho(t)=\dfrac{\text{e}^{-iH_{\mathcal{PT}}^{(anti)}}\rho(0)\text{e}^{iH^{(anti)\dagger}_{\mathcal{PT}}}}{\text{Tr}[\text{e}^{-iH_{\mathcal{PT}}^{(anti)}}\rho(0)\text{e}^{iH^{(anti)\dagger}_{\mathcal{PT}}}]},
 \end{equation}
 where the usual Hilbert-Schmidt inner product is employed. In this situation,    the effective  dynamics governed by $ H_{\mathcal{PT}}^{(anti)}  $ is    nonunitary and hence it describes  the evolution of an open quantum system \cite{kawabata2017information,ohlsson2020transition,jingwei2020observation}. It originates from the fact that a (an) (anti-)$ \mathcal{PT} \!-$symmetric system cannot be implemented by a closed system.  In other words, the lack of Hermiticity of the Hamiltonian,  is not observable in closed
 systems, in contrast to open systems \cite{brody2016consistency}. Hence, 
 we do not use a preferentially selected inner product
 with which the (anti-)$ \mathcal{PT} \!-$symmetric
 Hamiltonian $ H_{\mathcal{PT}}^{(anti)} $ can be 
 employed to generate the unitary time evolution for the characterization of a closed quantum system.
 The metric operators used in this approach for modifying the Hilbert space inner products are not physically observable \cite{brody2016consistency}, as discussed above.
 Such physical constraints prohibit experimentalists from modifying the inner product  in a laboratory, although it can be used as an effective mathematical tool to nicely formulate the theory of  quantum systems whose dynamics is governed by  (anti-)$ \mathcal{PT} \!-$symmetric
 Hamiltonians (see Ref.~\cite{ju2019non,dogra2021quantum} ).  
 
 \par
 
 In more detail, we see that the conventional  metric used in Eq.~(\ref{densityR}) leads to results completely consistent with experimental observations \cite{tang2016experimental,naghiloo2019quantum,bian2020quantum,jingwei2020observation,yu2020experimental,wang2020experimental,varma2021simulating,zhang2021observation,dogra2021quantum,wu2019observation} provided that it is applied correctly. In fact,  in order to physically implement the non-unitary evolution leading to Eq.~(\ref{densityR}), we can embed the $(anti)\mathcal{PT}\!-$symmetric system into a larger
 Hermitian system,  realized by adding
 ancillary qubits, and  perform a measuring process \cite{tang2016experimental,xiao2019observation}. This idea originates from the Naimark
 dilation procedure for quantum measurement \cite{naimark1943representation,hayashi2006quantum,kawabata2017information,huang2021solvable}: by including
 an ancilla  and extending the
 Hilbert space, any nonunitary dynamics can be implemented by
 a unitary dynamics of the total closed system followed by
 quantum measurement acting on the ancilla. When a measurement is performed on
 the ancilla and a special definite state is postselected,  the evolved state (\ref{densityR}) is realized.
 Because of this post-selection occurring in the  measuring  process, the successful implementation of the non-unitary gate is a probabilistic procedure. This experimental limitation \cite{dogra2021quantum,tang2016experimental,bian2020quantum,jingwei2020observation}, is similar to the situation which occurs in Bell inequality tests  \cite{hensen2015loophole,giustina2015significant,shalm2015strong}. Therefore, we  can solve the paradoxes \cite{lee2014local,bender2007faster,bender2013pt,ju2019non} associated with violation of \textit{no-go theorems} in $\mathcal{PT}\!-$symmetric theory   \cite{ju2019non} using normalized density matrix (\ref{densityR}),   without the need for any   modification in the Hilbert space (for more details, see Refs.~\cite{tang2016experimental,kawabata2017information}). 
 
 A significant property of the extended Hamiltonian discussed above
 is that the original system  is non-Hermitian if and only
 if the characteristic interaction  between the original system and  the ancilla is nonzero. On the one hand,  the information  flowed into the environment is actually stored in the entanglement  with the ancilla.
 On the other hand,  because the interaction is global, the
 quantum correlation  between
 the system and the ancilla   may oscillate in time \cite{kawabata2017information}.   Moreover, the information exchange between the system and this entangled partner hidden in the environment may be one of the
 physical origins of the time oscillations of the distance measures
 for quantum states  of the (anti-)$\mathcal{PT}\!-$symmetric
 system. The aforementioned reasons motivates us to define non-Markovianity concept in the (anti-)$\mathcal{PT}\!-$symmetric systems, however, as also described in the Introduction, the dependence of this definition on the postselection process, appearing in all current experimental realizations of (anti-)$\mathcal{PT}\!-$symmetric dynamics \cite{tang2016experimental,wu2019observation,naghiloo2019quantum,jingwei2020observation,bian2020quantum,yu2020experimental,wang2020experimental,dogra2021quantum}, may call into question its validity. Demonstrating the failure of the BLP measure in defining  possible non-Markovian effects in  (anti-)$\mathcal{PT}\!-$symmetric systems (see Refs. \cite{bian2020quantum,ding2021information}) can support this reasoning.

 \par  
 Adopting the BLP's definition of non-Markovianity, one finds that the \textit{P divisibility} of a linear quantum dynamical map is equivalent to Markovianity of the dynamics \cite{rivas2010entanglement}.  In spite of the fact that the (anti-)$ \mathcal{PT}$ dynamics given by Eq.~(\ref{densityR}) is indeed P divisible \cite{brody2012mixed}, the divisibility cannot capture the non-Markovianity of the (anti-)$ \mathcal{PT}$ dynamics because of the \textit{nonlinearity} of quantum operation
 $ \mathcal{E} $  \cite{kawabata2017information}. It is emphasized that even in the linear evolutions  the divisibility of map has  not be used to generally define  Markovianity, because the  concept of divisibility  is limited to processes for which
 the inverse of the dynamical map exists \cite{breuer2016colloquium,wissmann2015generalized}, a property which cannot be guaranteed e.g., examples on quantum semi-Markov
 processes \cite{vacchini2011markovianity} or the damped Jaynes-Cummings model on
 resonance \cite{laine2010measure},  thus making the concept of divisibility
 sometimes ill defined.
 \par
 In order to answer the question of whether the BLP's definition of non-Markovianity can be used in non-Hermitian systems \cite{kawabata2017information,xiao2019observation,jingwei2020observation}, we focus on TD contractivity.
 As known, the  TD always characterizes the distinguishability between two quantum states. When this property is associated with contractivity, it can be used for the definition of non-Markovianity interpreted as backflow of information from the environment to the system. However, in the absence of contractivity, its oscillation can only be attributed to oscillation of distinguishability, not existence of information backflow from the environment. Hence, to answer the question we should investigate the contractivity of TD in non-Hermitian systems.
 
 \par
 Concerning the mutual relations of the quantum operations and trace distance, as referred briefly in  Introduction, the following important result is well known \cite{nielsen2010quantum,rastegin2007trace,dajka2011distance,laine2011witness}: if there is no initial correlation between the system and environment and $ \mathcal{E} $ is a  trace-preserving quantum operation then 
 $
 D\big( \mathcal{E}(\rho), \mathcal{E}(\sigma)\big) \leq D\big( \rho, \sigma\big)
 $
 where $ \rho  $ and $  \sigma$ denote  arbitrary normalized quantum states. This result is usually referred to as contractivity
 of the trace distance under the linear trace-preserving quantum operations.
 \par
 However, 
 the quantum operation $ \mathcal{E} $, generating the evolved state (\ref{densityR}), is non-trace-preserving \cite{nielsen2010quantum}, since it does not provide a complete description of the processes
 occurring in the system. This nondeterministic feature originates from the fact that  other measurement outcomes may take place with some probability. Therefore, the contractivity of the TD under the evolution given in Eq. (\ref{densityR}) should be investigated in more detail.
 Our numerical calculation, presented in the next sections, shows that \textit{the trace distance  may exhibit non-contracticity under (anti-)$ \mathcal{PT}$ dynamics and hence in such systems it loses one of the necessary conditions which should be satisfied by a  faithful witness of non-Markovianity.}
 
 \par
 It should be noted  that contractivity is not a universal feature
 but depends on the metric: the dynamics may be
 contractive with respect to a given metric and may not be
 contractive with respect to other metric measures \cite{dajka2011distance}. In addition,
 contractivity of quantum evolution can break down when the system is initially correlated with its environment (for details see \cite{laine2011witness} ).

\section{Detecting criticality THROUGH HSS IN NON-HERMITIAN SYSTEMS}\label{sec:HSSnonHermitianMeasure}

In this section we provide the witness based on the Hilbert-Schmidt speed to faithfully identify the EPs and phase transitions 
in non-Hermitian systems. 

It is known that in the $ \mathcal{PT}$-symmetric systems the trace distance
oscillates with evolution time when the system symmetry is unbroken while in anti-$ \mathcal{PT}$-symmetric systems, the oscillations of the distinguishability occur if the  symmetry is broken  \cite{kawabata2017information,xiao2019observation,jingwei2020observation,dogra2021quantum}.  Inspired by this fact and the close relationship between HSS and TD described in Ref. \cite{jahromi2020witnessing}, we  propose the following easily computable witness to characterize the criticality in (anti-)$ \mathcal{PT}$-symmetric systems:


\textit{For a quantum system with an $n$-dimensional Hilbert space $\mathcal{H}$, let us consider an initial state given by 
	\begin{equation}\label{initialstate}
		|\psi_{0}\rangle=\dfrac{1}{\sqrt{n}}\big(\text{e}^{i\varphi}|\psi_{1}\rangle+\ldots+|\psi_{n}\rangle\big),
	\end{equation}
	where $\varphi$ is an unknown phase shift and $\{|\psi_{i}\rangle,\ i=1,\ldots,n\}$ represents the computational orthonormal basis. Then, we find  that   when the dynamics of the HSS, computed with respect to the initial phase $ \varphi $,  exhibit an  
	oscillating pattern, the (anti-)$\mathcal{PT}\!-$symmetric system is in (anti-)$\mathcal{PT}\!-$unbroken (broken) phase. 
	At EPs or broken (unbroken) phase of the (anti-)$\mathcal{PT}\!-$symmetric system, no oscillation is observed in the HSS dynamics.
	Therefore, the EPs can be easily detected by investigating the time evolution of the HSS.}



The sanity check of this protocol
as a faithful witness of EPs and criticality    is performed in the
following section.
\par


\section{$\mathcal{PT}$-symmetric two-level system}\label{SECTIONEXAMPLES}

\subsection{Hamiltonian model and computing the witness} 

As the first example, we  consider
the paradigmatic model of a two-level system described by the Hamiltonian 
\begin{equation}\label{HoneQubit}
	H_{\mathcal{PT}}=\varepsilon(\sigma_{x}+ia\sigma_{z})=\left(
	\begin{array}{cc}
		ia\varepsilon&\varepsilon \\
		\varepsilon&-ia\varepsilon \\
	\end{array}
	\right),
\end{equation}
where $ \varepsilon \geqslant 0 $ is an energy scale and $ a\geqslant 0 $ denotes the degree of non-Hermiticity. This model has been previously realized in both classical \cite{ruter2010observation,gao2015observation,liu2016metrology}
and quantum \cite{li2019observation,tang2016experimental,jingwei2020observation} experiments. The eigenenergies are  
$ \pm \varepsilon \sqrt{1-a^{2}} $,
and therefore one has an EP2 (exceptional point of order 2) given by $ a=1 $.
\par
The time evolution operator of this system is obtained as \cite{kawabata2017information}
\begin{eqnarray}\label{UoneQubit}
	& U_{\mathcal{PT}}=\text{e}^{-iH_{\mathcal{PT}}t}&\nonumber\\
	&=\dfrac{1}{\sqrt{1-a^{2}}}\left(
	\begin{array}{cc}
		\sqrt{1-a^{2}} \text{cos}\theta+a~ \text{sin}\theta & -i ~\text{sin}\theta \\
		-i~ \text{sin}\theta & \sqrt{1-a^{2}} \text{cos}\theta-a~ \text{sin}\theta \\
	\end{array}
	\right),&\nonumber \\
\end{eqnarray}
where $\theta =\sqrt{1-a^{2}}~\varepsilon t  $. 
\par
In order to check the efficiency of  our HSS-based witness, we first compute the normalized evolved state of the system when it is prepared in the initial state $\rho_{0}=\ket{\psi_{0}}\bra{\psi_{0}} $, where  $ |\psi_{0}\rangle=(\text{e}^{i\varphi}|0\rangle+|1\rangle)/\sqrt{2}$. We can easily calculate the HSS analytically by inserting the evolved state $ \rho_{t}(\varphi)=U_{\mathcal{PT}}\rho_{0}U_{\mathcal{PT}}^{\dagger}\bigg\{\text{Tr}[U_{\mathcal{PT}}\rho_{0}U_{\mathcal{PT}}^{\dagger}]\bigg\}^{-1} $ into Eq.~(\ref{HSSS}); however, its explicit expression has a cumbersome form and is not reported here.

\subsection{Dynamical behavior of the witness of the quantum criticality}

\begin{figure}[t!]
	\subfigure{ \includegraphics[width=0.45\textwidth]{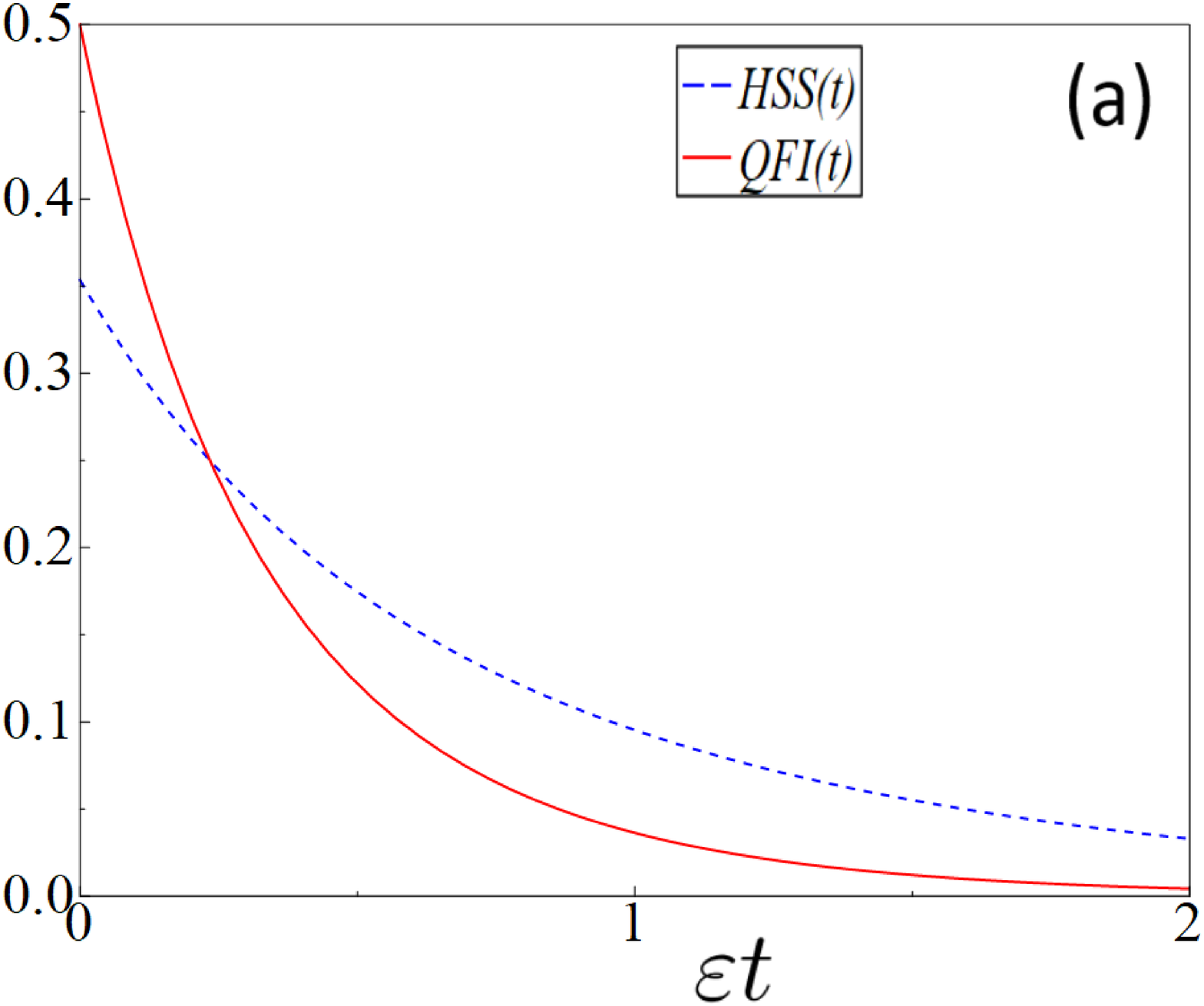}\label{HssQfi12} }
	\subfigure{\includegraphics[width=0.45\textwidth]{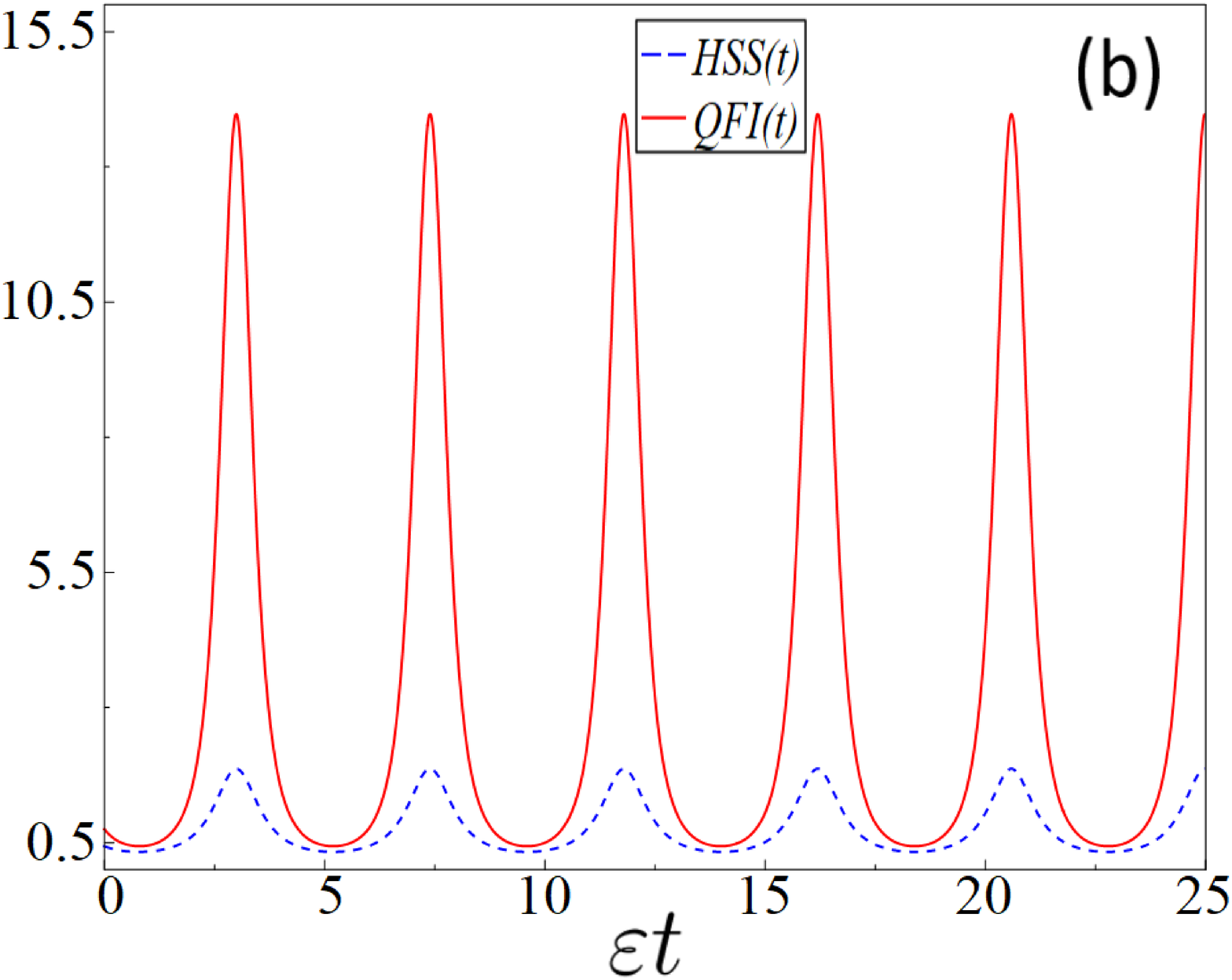}\label{HssQfi11}} 
	\caption{Dynamics of the Hilbert-Schmidt speed $ HS\!S(\rho_{t}(\varphi)) $ (blue
		dashed curve) and  the Quantum Fisher information $ QFI(t) $ (red solid curve) as a function of the dimensionless
		time $ \varepsilon t $ for a two-level system (a) in the $ \mathcal{PT} \!-$broken phase ($ a > 1 $) and (b) in the $ \mathcal{PT} \!-$unbroken phase ($ 0<a<1 $).}
	\label{HssQfi} 
\end{figure}

The qualitative dynamics of the HSS is displayed in Fig.~\ref{HssQfi} for the $ \mathcal{PT}  \!-$broken phase ($ a > 1 $) and $ \mathcal{PT}  \!-$unbroken phase ($ 0<a<1 $).
In the $ \mathcal{PT} \!-$broken phase,  the HSS, as expected, exhibits no oscillations and monotonically decreases with time. 
However, in the $ \mathcal{PT}  \!-$unbroken phase, it oscillates and eventually returns to its initial value, namely  there is a $ T $
\begin{equation}\label{InfRet}
	\ \text{such that}~~HS\!S(T)=HS\!S(0),
\end{equation}
where the period is given by $ T=\pi~ \big [1-a^{2}\big]^{-1/2} $. This period $ T $  of the
oscillation, called the recurrence time, increases as the system approaches the EP  ($ a=1 $).   This period is exactly similar to one  achieved theoretically as well as experimentally for the distinguishability oscillations \cite{kawabata2017information,xiao2019observation}. Moreover,  checking  the experimental data presented in \cite{xiao2019observation}, we find that the EP predicted by the HSS-based witness is quite accurate.

\subsubsection*{Behavior at the exceptional point} 

As discussed in the Introduction, the exceptional points (EPs) in non-Hermitian Hamiltonians constitute a threshold for the system parameters individuating a phase transition of the system. In the present case of a $ \mathcal{PT} \!-$symmetric system, one has a broken-to-unbroken phase transition by diminishing the value of the non-Hermiticity degree $a$ appearing in the Hamiltonian of Eq.~(\ref{HoneQubit}). Since this phase transition is associated to intriguing physical phenomena, the  characterization of the behavior of the non-Hermitian system around EPs is important. 
Moreover, because of the importance of the HSS in detecting the EPs, it would be interesting to investigate the behavior of this measure when the system approaches the EP, i.e., when $ a\rightarrow 1 $. In this limit, we find that the HSS  is given by
\begin{eqnarray}
	HS\!S_{EP}(t)&=&\dfrac{1}{  4\,{\varepsilon}^{2}{t}^
		{2}-4\,{\varepsilon}^{2}{t}^{2}\sin \left( \varphi  \right)+2}.
\end{eqnarray}
It is immediate to see that it decreases monotonically with time according to  $HS\!S_{EP}(t)\propto t^{-2} $; while only for $ \varphi=\pi/2 $ the HSS gives a constant value equal to $ 1/2 $. These results immediately show that, when the system parameters reach the EP ($ a\rightarrow 1 $), no oscillation in HSS dynamics is observed which arises from the divergence of  period $ T $  at this limit.

Overall, we see that the HSS can be employed as an efficient witness to  identify phase transitions and detect the EPs in the non-Hermitian system under consideration.

\subsection{Relationship between quantum Fisher information (QFI) and HSS }\label{QFIHSS}

The fundamental question in the theory of quantum estimation is the following: When performing
measurements on the quantum systems affected by some classical parameter $ \varphi $ (which may be the phase encoded into the initial state of the system), how precisely
can $ \varphi $ be estimated? The answer is given by the quantum
Cramer-Rao bound \cite{braunstein1994statistical} indicating  that the smallest
resolvable change in $ \varphi $ is $ \delta \varphi =1/\sqrt{F_{\varphi}}  $ where $ F_{\varphi} $ denotes the quantum Fisher information (QFI)  given by \cite{braunstein1994statistical,liu2019quantum}
\begin{equation}
	F_{\varphi}(\rho(t))=2\sum_{i,j} \dfrac{|\langle \phi_{i}|\partial_{\varphi}\rho\left(\varphi \right)|\phi_{j}\rangle|^{2}}{(\lambda_{i}+\lambda_{j})},
\end{equation}
where $ |\phi_{i}\rangle $ and $ \lambda_{i} $ represent, respectively,  the eigenvectors and eigenvalues of  the density matrix $ \rho\left(t \right) $. It should be noted that recently other expressions for the QFI have been proposed in non-Hermitian systems \cite{li2021cram}, however, for pure states they reduce to the above expression, ignoring a constant coefficient. 

\par
Because both QFI and HSS are quantum statistical speeds associated, respectively, with the \textit{Bures} and \textit{Hilbert–Schmidt distances} (for details see \cite{gessner2018statistical}), it is reasonable to explore how they can be 
related to each other. 
Recently,  a strong relationship between the HSS and QFI
has been constructed in the process of phase estimation for $n$-qubit Hermitian systems  \cite{jahromi2021hilbert}. It has been found  that, when both the HSS and 
QFI are computed with respect to the phase parameter encoded into 
the initial state of an $n$-qubit system, the zeros of the HSS dynamics are actually equal to those of the QFI dynamics. Likewise, the signs of the time-derivatives of both HSS and QFI exactly coincide.
\par
Now computing the QFI with respect to the phase parameter encoded into the initial state of our one-qubit $ \mathcal{PT}  \!-$symmetric system, we obtain the similar result, i.e., both the QFI and HSS exhibit the same qualitative dynamics (see Fig. \ref{HssQfi}). Moreover,  in the $ \mathcal{PT} \!-$broken phase ($ a>1 $) illustrated in Fig.~\ref{HssQfi12}, both measures are contractive and hence monotonically decrease with time. However, as observed in Fig.~\ref{HssQfi11},
not only the HSS but also  the QFI,   contractive under CPTP maps in Hermitian systems, may  be non-contractive under the non-trace-preserving evolution in the unbroken phase ($ 0<a<1 $). Moreover, this important relationship between the HSS and QFI shows that the HSS may be introduced as an efficient figure of merit for quantum estimation of phase encoded into the initial state of $n$-qubit $ \mathcal{PT}  \!-$symmetric systems. This will be investigated in detail in future studies. 

The non-contractivity of the QFI and HSS confirms the fact that quantum information measures and witnesses may exhibit different behaviors for Hermitian and non-Hermitian systems. 
It is known that the QFI measures the maximum information about a parameter $ \varphi $, extractable from a given measurement procedure \cite{braunstein1994statistical,liu2019quantum,petz2011introduction,jafarzadeh2020effects,shadehi2020adiabatic}. 
In our model, because the parameter to be estimated is encoded into the initial state of the system and the system, initially not correlated with the environment, does not sense it later, 
we reasonably expect that the maximum information, achieved in the estimation process, must be extracted from the initial state itself.   
Contrary to this intuitive reasoning, it is observed that the non-Hermitian evolution of the system interestingly may enhance the estimation of the initial parameter with time. 
\par
We can conclude that the system initially hides some of the encoded information such that it is inaccessible at first and then the non-Hermitian evolution makes this hidden information available for the estimation of the initial phase. In fact, this unusual behavior, leading to non-contractivity of both QFI and HSS, is indistinguishable from the non-Markovian behavior (backflow of information from the environment to the system). 
This is why we cannot use the standard Hermitian witnesses based on QFI and HSS to detect the non-Markovianity in this context.

\subsection{Non-contractivity of TD  in  $\mathcal{PT}\!-$symmetric one-qubit systems}
\begin{figure}[t!]
	\includegraphics[width=0.45\textwidth]{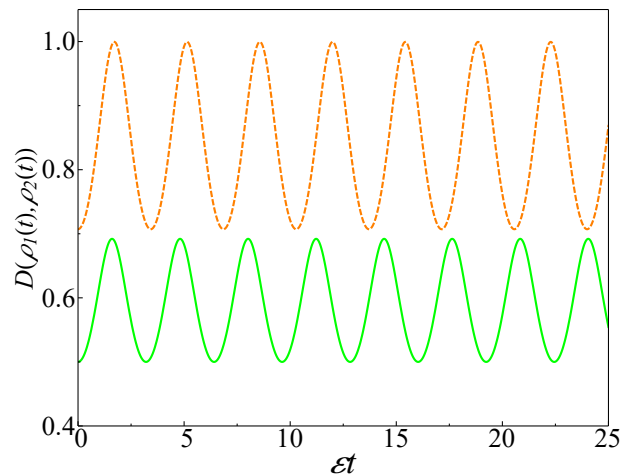}
	\caption{Dynamics of the distinguishability $D $ between two evolved states $ \rho_{1} (t)$ and $ \rho_{2} (t)$ starting from two special pairs of initial states as a function of the dimensionless
		time $ \varepsilon t $ for a two-level system  in the $ \mathcal{PT} \!-$unbroken phase ($ 0<a<1 $).  }
	\label{noncont1} 
\end{figure}
The non-contractivity of the QFI and HSS motivates us to investigate the TD non-contractivity under non-Hermitian evolution of the system. Our numerical computation shows that the TD may be non-contractive in such systems. Assuming two initial states $ \ket{\psi _{0}^{(1)}}=1/\sqrt{2}\big(\text{e}^{i\varphi} \ket{0}+\ket{1}\big) $ and $  \ket{\psi _{0}^{(2)}}=1/\sqrt{2}\big( \ket{0}-\text{e}^{i\varphi}\ket{1}\big) $, and computing the TD between the two corresponding evolved states, we can observe the non-contractivity of the TD under $ \mathcal{PT}$ dynamics (see Fig. \ref{noncont1}, showing the time evolution of  the TD, in which the green solid and orange dashed curves are plotted for ($ a=0.2, \varphi=\pi/3 $ ) and ($ a=0.4, \varphi=\pi/4 $), respectively).

In our model, initially the system is  not correlated with the environment. This preparation alongside the fact that the decoherence effects lead to flow of information from the system to the environment, make us
reasonably expect that the  distinguishability shows contractivity under the time evolution. However, as discussed in Sec.~\ref{QFIHSS}, the system initially may hide some of the information such that it is inaccessible at first and then the non-Hermitian evolution makes this hidden information available for the system. 
This counterintuitive behavior may be one of the reasons that the TD exhibits non-contractive behavior. Moreover, this unusual behavior is indistinguishable from the non-Markovian behavior (i.e., backflow of information from the environment to the system). Therefore,  we cannot use the BLP's measure as a faithful witness  to detect the non-Markovianity in $\mathcal{PT}\!-$symmetric systems.

\section{ ANTI-$ \mathcal{PT}  \!-$SYMMETRIC TWO-LEVEL SYSTEM}\label{ANTISECTIONEXAMPLES}

\subsection{Hamiltonian model and witnesses} 

The generalized form of a single-qubit anti-$\mathcal{PT}\!-$symmetric Hamiltonian can be expressed as \cite{bender2002complex}
\begin{equation}\label{HantiPT}
	H^{\text{anti}}_{\mathcal{PT}}=\left(
	\begin{array}{cc}
		\lambda \eta ~\text{e}^{i\vartheta} &i\eta \\
		i \eta&-\lambda \eta ~\text{e}^{-i\vartheta} \\
	\end{array}
	\right),
\end{equation}
where all of the parameters $ \lambda, \vartheta $ and $ \eta$ denote real numbers. It is easy to show that this Hamiltonian satisfies   the anti-commutation relation
\begin{equation}
	(\mathcal{PT})H^{\text{anti}}_{\mathcal{PT}}(\mathcal{PT})^{-1}=-(H^{\text{anti}}_{\mathcal{PT}})^{T}=-H^{\text{anti}}_{\mathcal{PT}}
\end{equation}
where here operator $\mathcal{P}$ denotes Pauli matrix $ \sigma_{x} $, $\mathcal{T}$ represents the  complex
conjugation, and   notation $ A^{T} $ means the transpose of matrix $ A $. 
The
eigenvalues of Hamiltonian $ H^{\text{anti}}_{PT} $ are given by $ \epsilon_{\pm}=i\lambda \eta~\text{sin}\vartheta \pm \sqrt{\lambda^{2}\eta^{2}\text{cos}^{2}\vartheta-\eta^{2}} $ and the system is denoted in the regime of unbroken anti-$\mathcal{PT}\!-$symmetric phase if $ \lambda^{2}\eta^{2}\text{cos}^{2}\vartheta-\eta^{2}<0 $. For simplicity, as well as comparison with the experimental results presented in \cite{jingwei2020observation} for this kind of system, we consider the scenario in which $ \vartheta=0 $, $ \eta \geq 0 $ being  an energy scale and $ \lambda \geq 0 $ denoting the degree of Hermiticity. Therefore,  the EP2 is located at $ \lambda=1 $.
  We obtain the corresponding time evolution
operator of this system as 
\begin{equation}
	U^{\text{anti}}_{\mathcal{PT}}=\left(
	\begin{array}{cc}
		\cos  \theta-\frac{i \lambda \sin  \theta}{\sqrt{\lambda^2-1}} & \frac{\sin  \theta}{\sqrt{\lambda^2-1}} \\
		\frac{\sin \theta}{\sqrt{\lambda^2-1}} & \cos  \theta+\frac{i \lambda \sin  \theta}{\sqrt{\lambda^2-1}} \\
	\end{array}
	\right),
\end{equation}
where $ \theta=\sqrt{\lambda^2-1}~ \eta t $. 
\par

The calculation of the HSS  is similar to the approach followed in the previous section. The analytical expressions for this witness is also accessible; however, it does not have  compact informative forms, and hence it is not reported here. 

\subsection{Dynamical behavior of the witness of quantum criticality}
\begin{figure}[t!]
	\subfigure{ 	\includegraphics[width=0.435\textwidth]{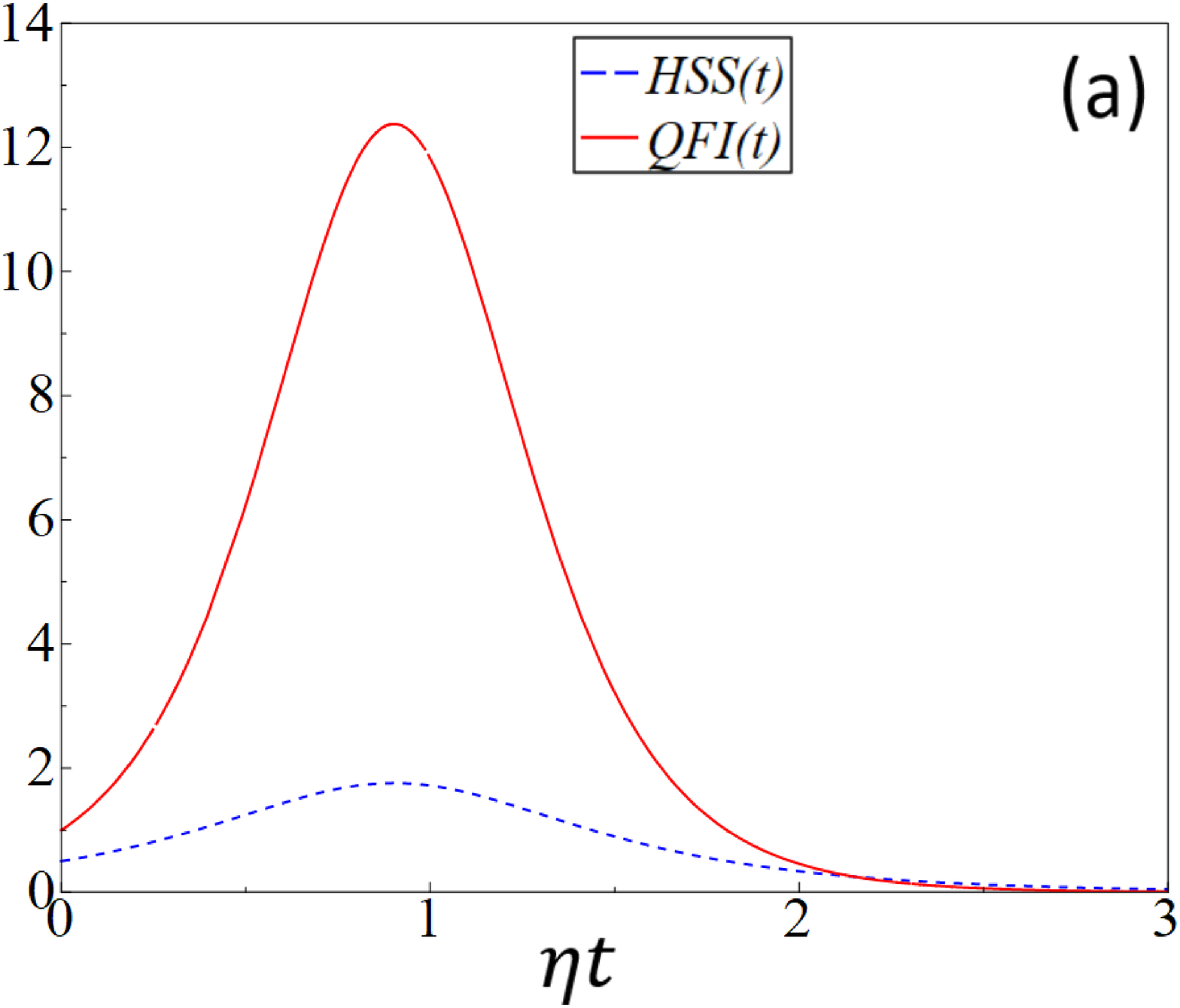}\label{HssQfianti21} }
	\subfigure{  \includegraphics[width=0.435\textwidth]{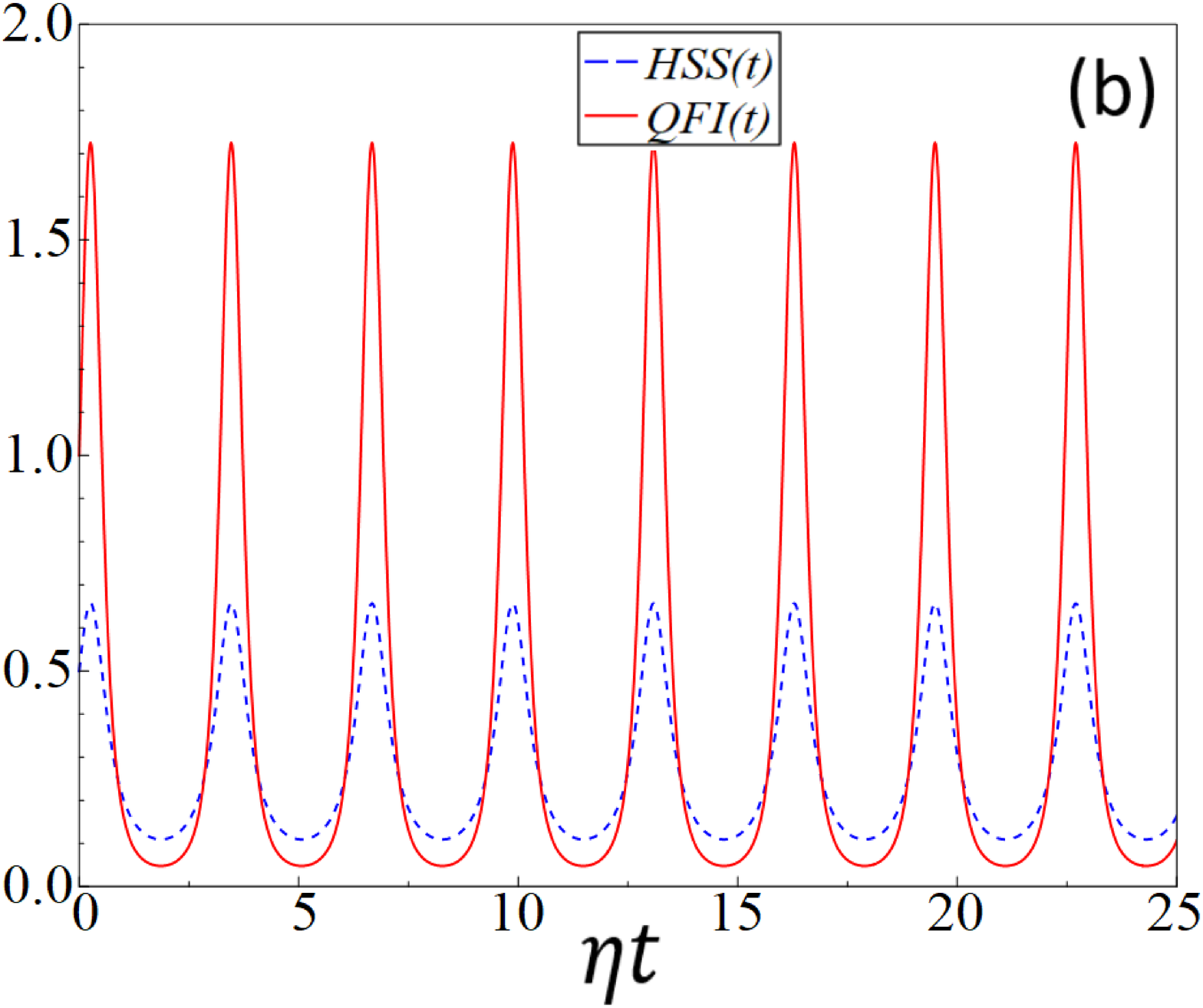}\label{HssQfianti22} }
	\caption{Dynamics of the Hilbert-Schmidt speed $ HS\!S(\rho_{t}(\varphi)) $ (blue
		dashed curve) and QFI $ D(t) $ (red solid curve) as a function of the dimensionless
		time $ \eta t $  for a two-level system (a) in the anti-$ \mathcal{PT}  \!-$unbroken phase ($ \lambda < 1 $) and (b) in the anti-$ \mathcal{PT}  \!-$broken phase ($ \lambda >1 $).}
	\label{HssQfianti} 
\end{figure}
We again see that the HSS works well in detecting the critical behavior of the system. In the anti-$ \mathcal{PT}  \!-$unbroken phase ($ \lambda < 1 $), as shown in Fig.~\ref{HssQfianti21} plotted for $ \lambda=0.2 $, the HSS does not oscillate over time. However,   its oscillatory behavior for $ \lambda >1 $  exactly predicts that the system is in anti-$ \mathcal{PT} \!-$broken phase (see Fig.~\ref{HssQfianti22} plotted for $ \lambda =1.4 $).
Moreover, we find  that the intensity of the HSS  oscillations is related to the parameter $ \lambda $, denoting the degree of Hermiticity, such that with an increase in this parameter, its oscillation gradually weakens.

In  Ref.~\cite{jingwei2020observation}, the authors
proposed an algorithm for the implementation of
the above generalized anti-$\mathcal{PT}\!-$symmetric evolution with a circuit-based
quantum computing system applying a three-qubit scheme including two ancillary
qubits and one working qubit. The
implementation scheme is based on decomposing the non-Hermitian
Hamiltonian evolution into a linear combination of unitary
operators and realizing the scheme in an enlarged Hilbert
space with ancillary qubits. 
The experimental results     clearly show that the distinguishability
oscillates with period
\begin{equation}
	T=\dfrac{\pi}{\eta \sqrt{\lambda^{2}-1}},
\end{equation}
when the system symmetry is broken. Moreover, the intensity of the distinguishability oscillation is  connected to  parameter $ \lambda $ such that the transition between the broken and unbroken phases is completely determined by this parameter. Comparing our findings to these experimental results, we see that they are completely consistent. In particular we find that the periods of the HSS and distinguishability oscillations are exactly the same.
This fact hence proves the efficiency of our proposed witness in faithfully detecting the critical behavior of the anti-$ \mathcal{PT}  \!-$symmetric systems.

\subsubsection*{Behavior at the exceptional point}

In this case of an anti-$\mathcal{PT}\!-$symmetric system, the exceptional point (EP) individuates a phase transition of the system from the unbroken to the broken phase when increasing the parameter $\lambda$ of the Hamiltonian of Eq.~(\ref{HantiPT}).
In order to complete the analysis, we investigate the behavior of the HSS  at the EP  $ \lambda=1 $, where it is given by
\begin{eqnarray}
	HS\!S_{EP}^{\text{anti}}(t)&=&\frac{1}{2 \left| 2 \eta^2 t^2+2 \eta t \left[ \eta t~ \sin  \varphi +\cos  \varphi \right]   +1 \right|  }.
\end{eqnarray}
It is easily seen that the  time behavior of the HSS is similar to that we have shown in Fig.~\ref{HssQfianti21} depicting the HSS dynamics in the unbroken phase.  This result also confirms that  the HSS can be used to detect phase transitions at the EPs in this type of non-Hermitian systems.

\subsection{Relationship between QFI and HSS }

Calculating  the QFI with respect to  phase parameter
$ \varphi $, we again find that both the QFI and HSS exhibit the same qualitative dynamics in this one-qubit anti-$ \mathcal{PT}  \!-$symmetric system (see Fig.~\ref{HssQfianti}). 

Interestingly,   in both  unbroken and broken  phases, depicted in  Fig.~\ref{HssQfianti}, both measures does not  necessarily exhibit contractivity, usually assumed as a necessary property for any faithful witness of non-Markovianity. 

\subsection{Non-contractivity of TD  in  anti-$\mathcal{PT}\!-$symmetric one-qubit systems}
\begin{figure}[t!]
	\subfigure{ 	\includegraphics[width=0.435\textwidth]{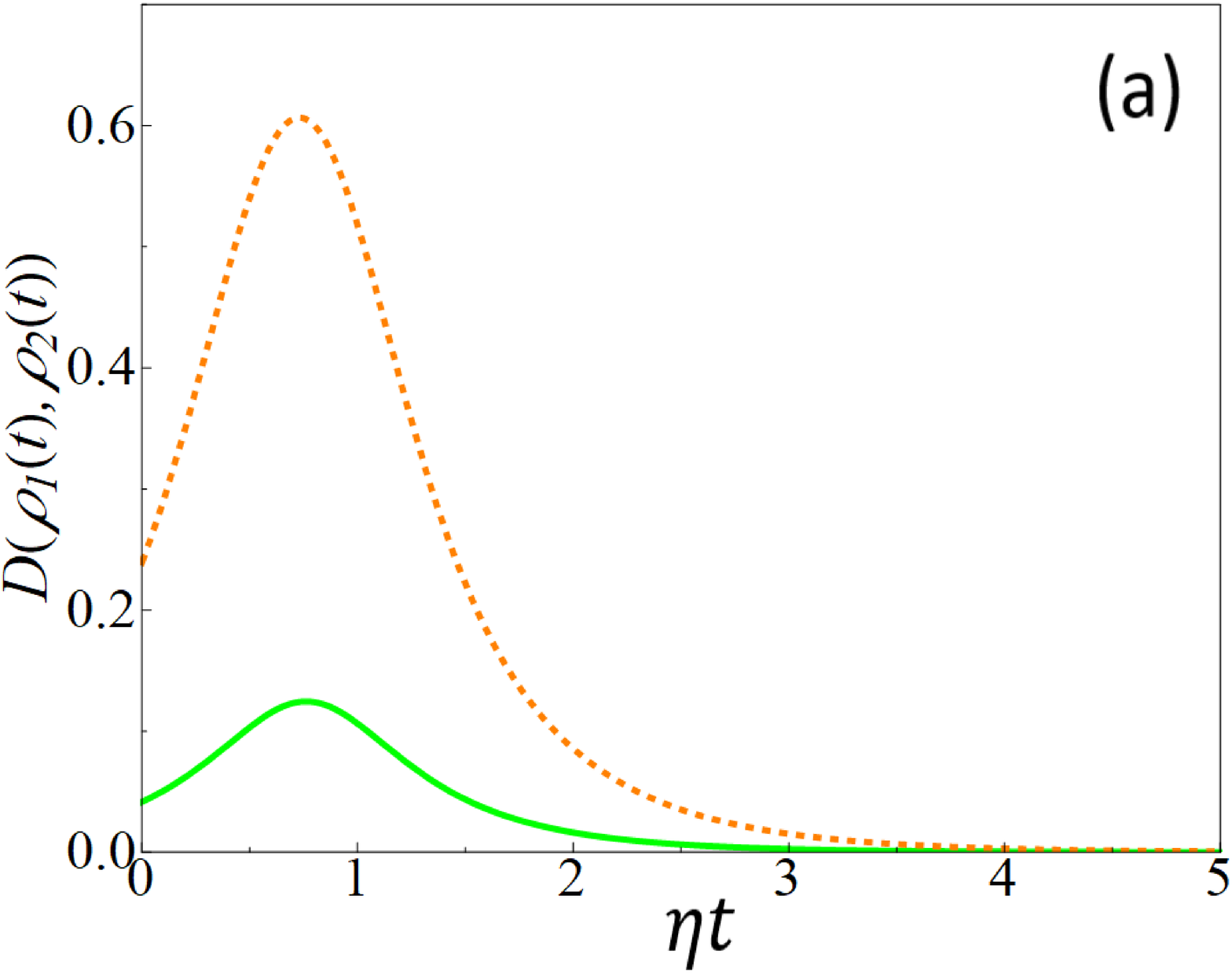}\label{antinoncon1} }
	\subfigure{  \includegraphics[width=0.435\textwidth]{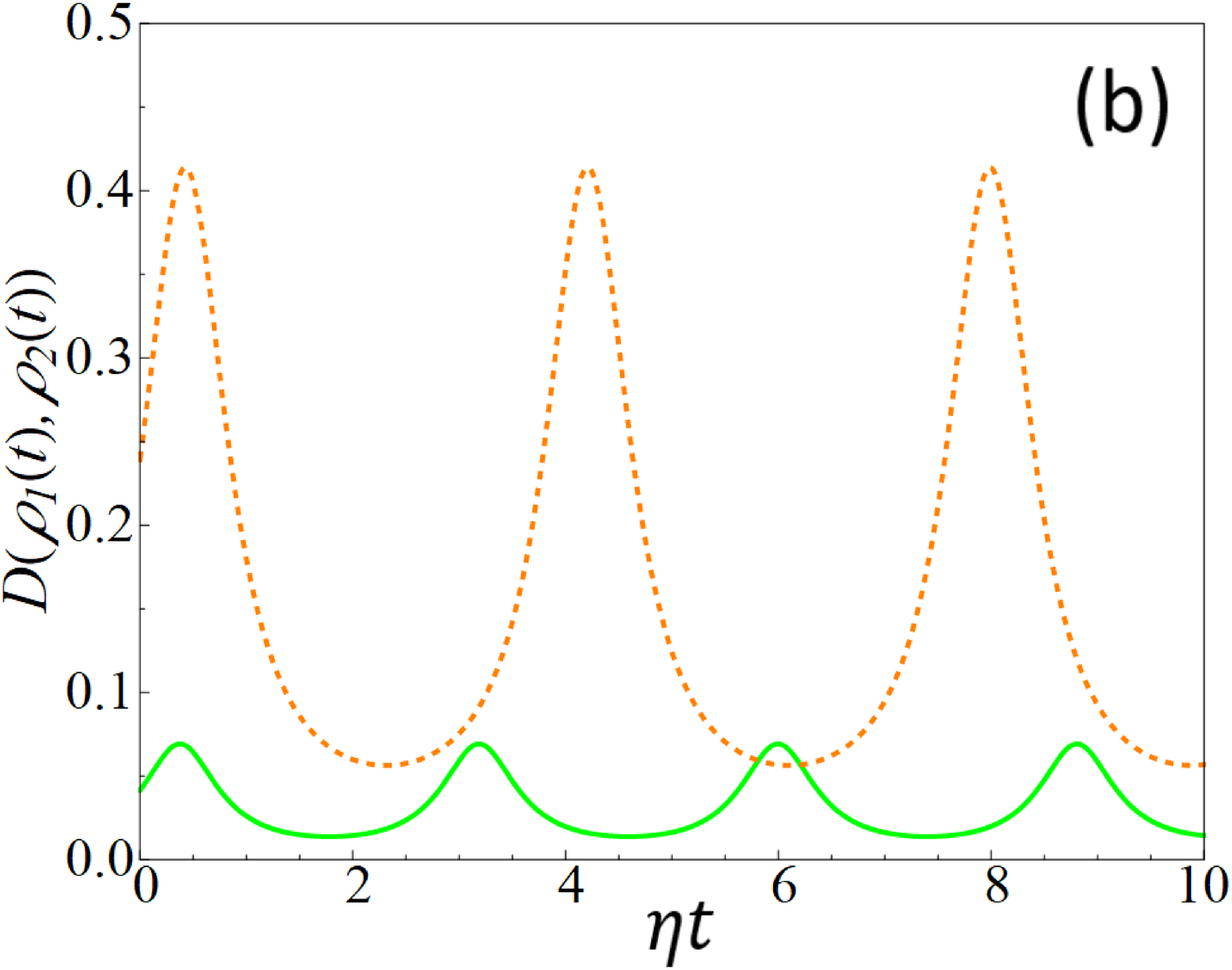}\label{antinoncon2} }
	\caption{Dynamics of the distinguishability $D $ between two evolved states $ \rho_{1} (t)$ and $ \rho_{2} (t)$ starting from two special pairs of initial states as a function of the dimensionless
		time $ \eta t $  for a two-level system (a) in the anti-$ \mathcal{PT}  \!-$unbroken phase ($ \lambda < 1 $) and (b) in the anti-$ \mathcal{PT}  \!-$broken phase ($ \lambda >1 $). }
	\label{antinoncon} 
\end{figure}

Numerically investigating  the TD contractivity under the anti-$\mathcal{PT}\!-$symmetric evolution of the system, we find  that it is not necessarily  contractive in these systems. Starting from two initial states $ \ket{\psi _{0}^{(1)}}=1/\sqrt{2}\big(\text{e}^{i\varphi} \ket{0}+\ket{1}\big) $ and $  \ket{\psi _{0}^{(2)}}=1/\sqrt{2}\big( \ket{0}+\text{e}^{i\varphi}\ket{1}\big) $, and computing the TD between the two corresponding evolved states, we may observe the non-contractivity of the TD under anti-$ \mathcal{PT}$ symmetric dynamics in both unbroken and broken phases (see Fig. \ref{antinoncon1} (\ref{antinoncon2}), showing the time evolution of  the TD in the unbroken (broken) phase, where the green solid and orange dashed curves are plotted for $ \bigg(\lambda=0.5 ~(\lambda=1.5), \varphi=3.1 \bigg)$  and $ \bigg(\lambda=0.6 ~(\lambda=1.3), \varphi=2.9 \bigg)$, respectively).


  \section{ HIGH-DIMENSIONAL $ \mathcal{PT}\!-$SYMMETRIC SYSTEM}\label{qudit}
 
 \subsection{Hamiltonian model and witnesses} 
 
 Now we consider an open, high-dimensional system described by a $4 \times 4$ Hamiltonian \cite{bian2020quantum}
 \begin{equation}
 	H_{\mathcal{PT}}=-JS_{x}+i\gamma S_{z}
 \end{equation}
 in which $S_{x}  $ and $ S_{z} $ denote spin-3/2 representations of the SU(2)
 group. In the orthonormal  computational basis $ \{\ket{1},\ket{2},\ket{3},\ket{4}\} $, the Hamiltonian can be written in the following  matrix form 
 \begin{equation}\label{HQubit}
 	H_{\mathcal{PT}}=\dfrac{1}{2}\left(
 	\begin{array}{cccc}
 		3i\gamma &-\sqrt{3}J&0 &0 \\
 		-\sqrt{3}J &i\gamma&-2J &0 \\
 		0 &-2J &-i\gamma&-\sqrt{3}J \\
 		0 &0&-\sqrt{3}J &-3i\gamma \\
 	\end{array}
 	\right),
 \end{equation}
 representing a $ \mathcal{PT}\!-$symmetric qudit with $ d=4 $.  The eigenvalues of  $ H_{PT} $ are
 simply given by $ \lambda_{k}=\{-3/2,-1/2,1/2,3/2\}\sqrt{J^{2}-\gamma^{2}}$ $(k=1,2,3,4) $,  leading to an EP4 at the $ \mathcal{PT}\!-$breaking
 threshold $ \gamma=J $. This Hamiltonian can be easily generalized to an arbitrary dimensional
 system and has an EP with the order equal to the dimension of the system  \cite{hodaei2017enhanced,graefe2008non,quiroz2019exceptional}.
 \begin{figure}[t!]
 	\centering
 	\subfigure{ \includegraphics[width=0.48\textwidth]{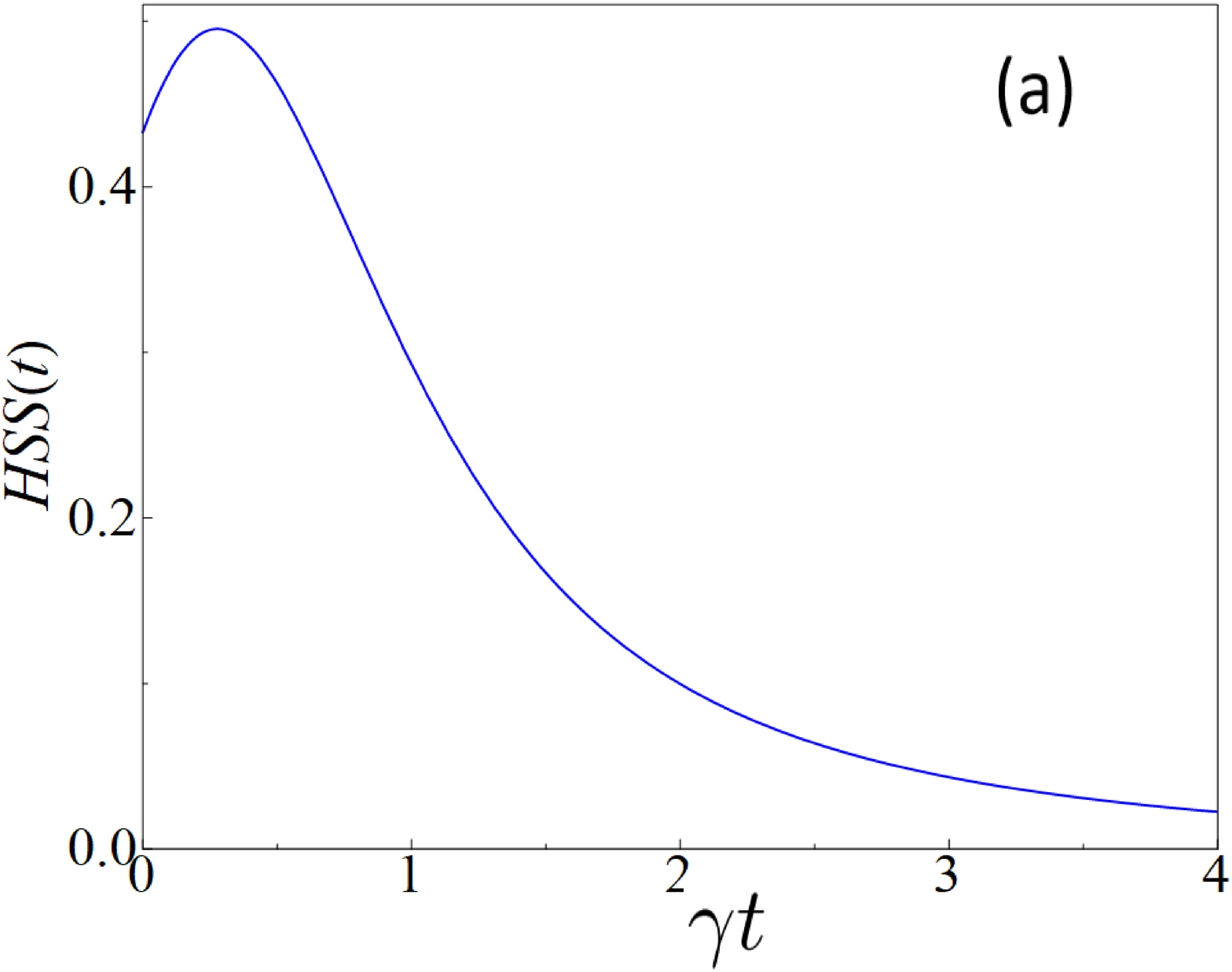}\label{HSSqudit1} }
 	\subfigure{ 	\includegraphics[width=0.48\textwidth]{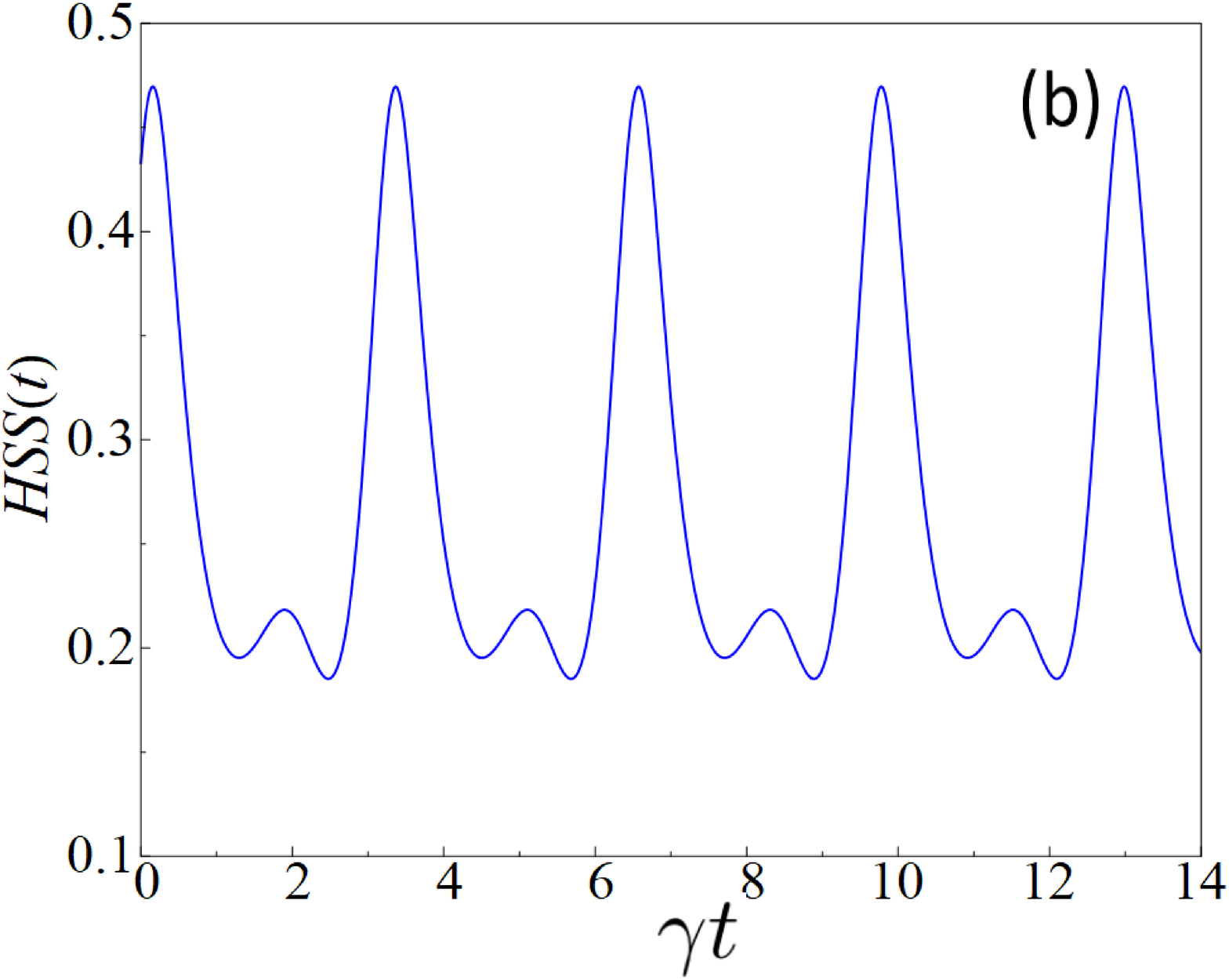}\label{HSSqudit2} }
 	\caption{Dynamics of the Hilbert-Schmidt speed $ HS\!S(\rho_{t}(\varphi)) $ as a function of the dimensionless
 		time $ \gamma t $ for the qudit system (a) in the $ \mathcal{PT}\!-$broken phase ($ \gamma >J $) and (b) in the $ \mathcal{PT} \!-$unbroken phase ($ \gamma<J $).  
 	}
 	\label{HSSqudit} 
 \end{figure}

 Let us now assume that the qudit system is $4$-dimensional ($d=4$) and prepared in the pure initial state of Eq.~(\ref{initialstate}): $ \ket{\psi_{0}}=\big(\text{e}^{i\varphi}\ket{1}+\ket{2}+\ket{3}+\ket{4}\big)/\sqrt{4} $.
 To obtain the time evolution operator $ U=\text{e}^{-iH_{\mathcal{PT}}t} $, we can expand the initial state  in terms of  the non-orthogonal  eigenvectors $ \ket{\zeta_{k}} $  of the  non-Hermitian Hamiltonian $ H_{\mathcal{PT}} $ as $ \ket{\psi_{0}}=\sum_{k}^{}\beta_{k}\ket{\zeta_{k}}$, where the coefficients $\beta_{k}  $'s should be determined. To this aim, we first define a square $ d \times d $ matrix $ \Phi $ such that the normalized eigenvectors $ \ket{\zeta_{k}} $ are concatenated as its columns.  Therefore, the initial state can be represented as $\ket{\psi_{0}}:= \Phi~ \beta $
 where $ \beta$ denotes a column matrix with elements $  \beta_{i}$.  Because the  columns of $ \Phi $ are linearly independent, it is invertible, and hence we can write $\ket{\psi_{0}}:=\Phi \Phi ^{-1} \ket{\psi_{0}} $, leading to the relation $ \beta=\Phi ^{-1} \ket{\psi_{0}} $. After computing the coefficients $\beta_{k}  $'s by the above method, we can easily obtain the evolved state of the system. Although the computation of the HSS   (associated with the  evolved density matrix $ \rho_{t}(\varphi)=\ket{\psi(t)}\bra{\psi(t)} /\text{Tr}[\ket{\psi(t)}\bra{\psi(t)}]$ where $ \ket{\psi(t)}=\text{e}^{-iH_{\mathcal{PT}}t}  \ket{\psi_{0}}=  \sum_{k}^{}\text{e}^{-i\lambda_{k}t}\beta_{k}\ket{\zeta_{k}}$, with respect to initial phase $ \varphi $) is straightforward, the explicit analytic expression is extremely complex and is not reported in this paper, whereas the results
 are described below.

 \subsection{Dynamical behavior of the witness and EP role}
 
 Analyzing the HSS dynamics reveals that the EP4 with $ \gamma = J $ determines the border between $ \mathcal{PT}  \!-$symmetric unbrken and broken phases of the qudit (see Fig.~\ref{HSSqudit}). In the $ \mathcal{PT}\!-$broken phase ($ \gamma >J $), the HSS first may increase with time and show a peak. Nevertheless, no oscillation in its dynamics is observed and  then it monotonously falls with time  (see Fig. \ref{HSSqudit1} plotted for $ J/\gamma=0.9 $). However, in the $ \mathcal{PT}\!-$symmetry unbroken region ($ \gamma < J $), as expected and  shown in Fig. \ref{HSSqudit2} plotted for $ J/\gamma=2.2 $, the HSS dynamics exhibits periodic oscillations (see relation (\ref{InfRet})).

 \subsection{Non-contractivity of TD  in  high dimensional $\mathcal{PT}\!-$symmetric  systems}
 \begin{figure}[t!]
 	\centering
 	\subfigure{ \includegraphics[width=0.48\textwidth]{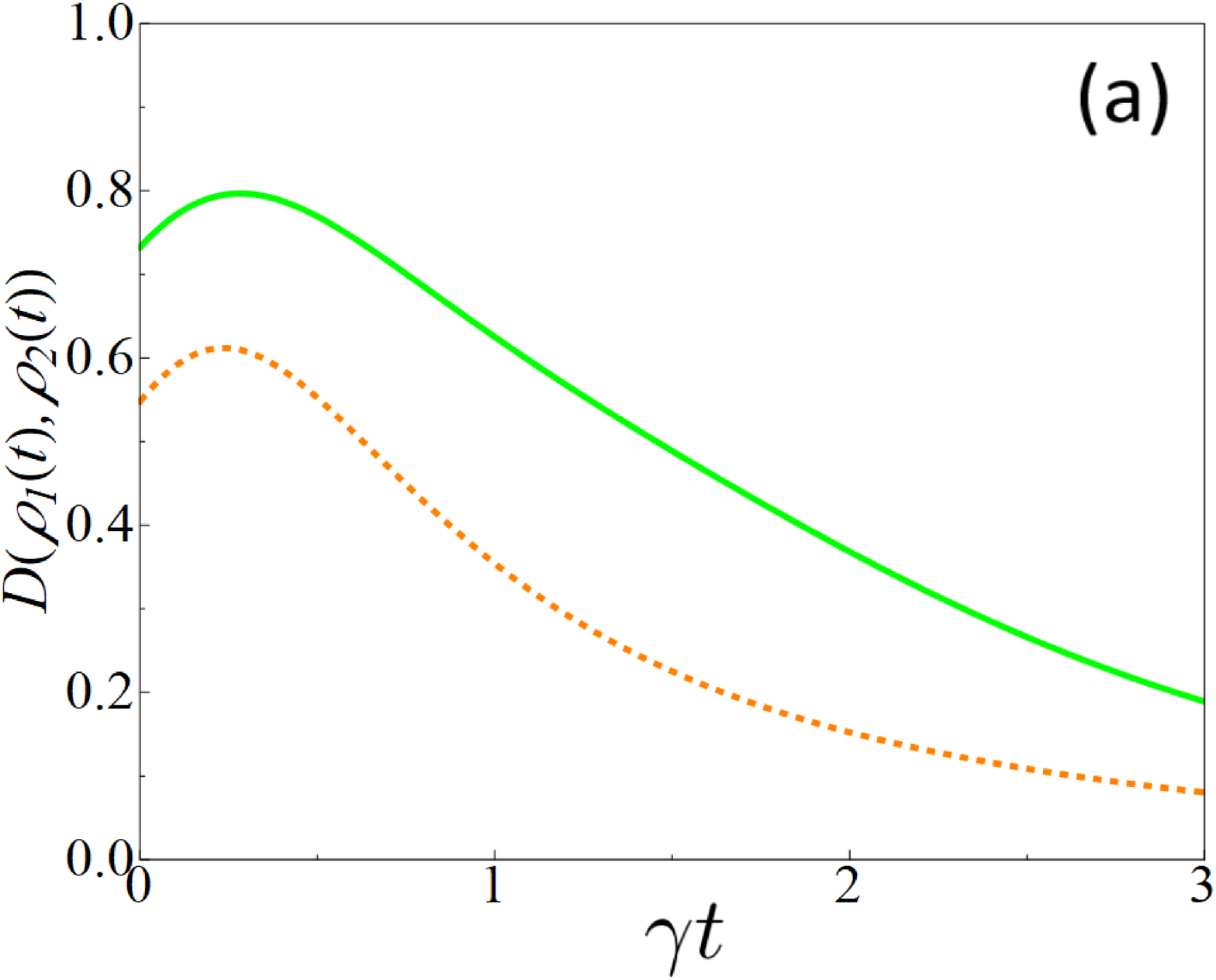}\label{qudit1} }
 	\subfigure{ 	\includegraphics[width=0.48\textwidth]{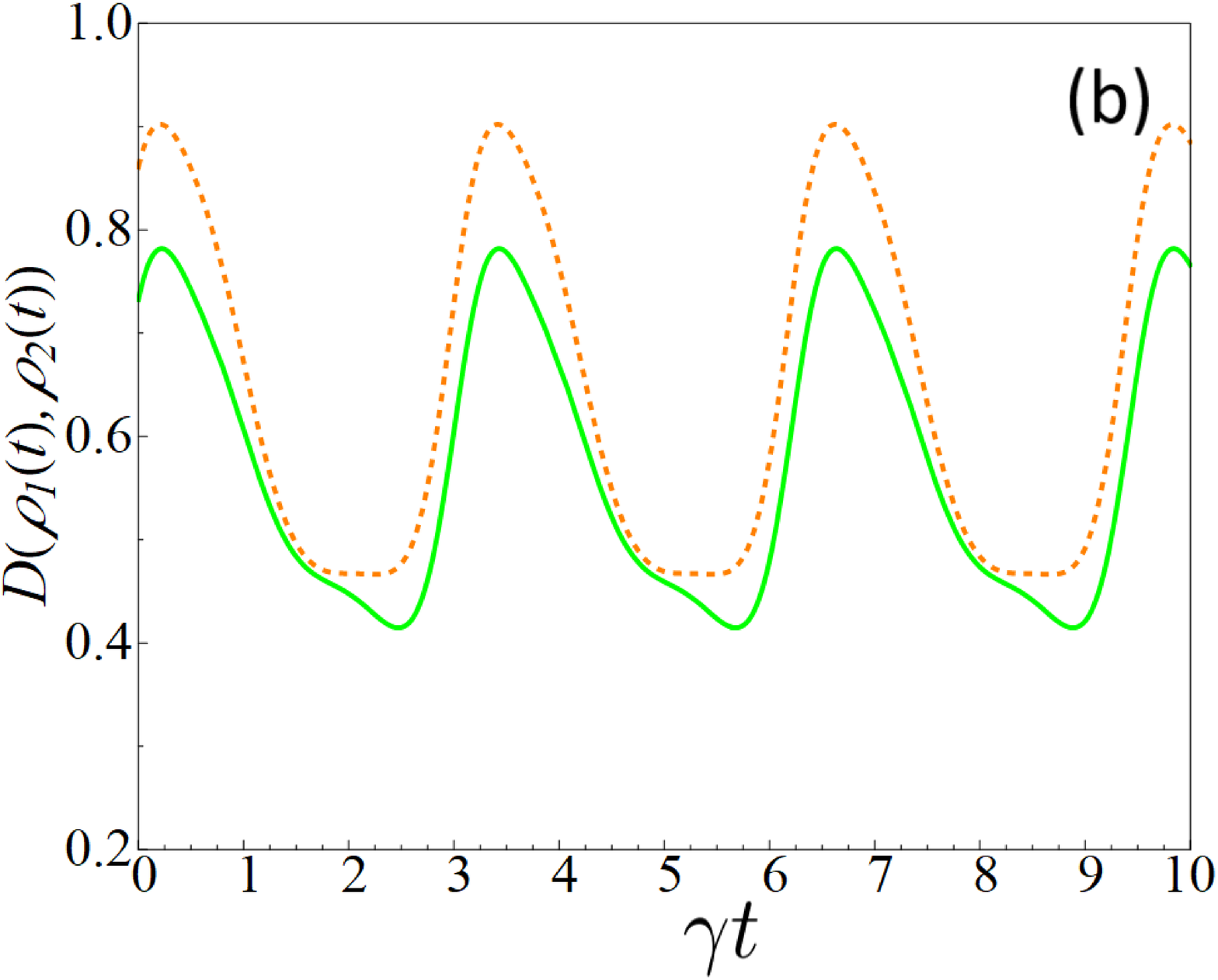}\label{qudit2} }
 	\caption{Dynamics of the distinguishability $D $ between two evolved states $ \rho_{1} (t)$ and $ \rho_{2} (t)$ starting from two special pairs of initial states of the qudit system as a function of the dimensionless
 		time $ \gamma t $   (a) in the $ \mathcal{PT}\!-$broken phase ($ J/\gamma=0.9 $) and (b) in the $ \mathcal{PT} \!-$unbroken phase ($ J/\gamma=2.2 $). 
 	}
 	\label{qudit12} 
 \end{figure}
 
 The computation of the trace distance requires diagonalization of   $ \rho_{1}(t) -\rho_{2}(t) $ for the pair of optimal initial states $ \rho_{1}(0) $ and $ \rho_{2}(0) $. The optimization process  with complexity of  computing  the eigenvectors of  the high dimensional complicated density matrices  $ \rho_{i}(t) ~ (i=1,2) $, makes the trace distance too difficult to compute.

 We check numerically   the  contractivity of TD under the $\mathcal{PT}\!-$symmetric evolution of the qudit for a large number of random pairs of initial states. We explicitly show   that the TD may exhibit non-contractivity in both unbroken and broken phases. For example, starting from two initial states $ \ket{\psi _{0}^{(1)}}=\big(\text{e}^{i\varphi}\ket{1}+\ket{2}+\ket{3}+\ket{4}\big)/\sqrt{4} $ and $ \ket{\psi _{0}^{(2)}}=\big(\text{e}^{i\theta}\ket{1}+\text{e}^{i\varphi}\ket{2}+\ket{3}+\ket{4}\big)/\sqrt{4} $, and computing the TD between the two corresponding evolved states, one can observe the non-contractivity of the TD under $ \mathcal{PT}$ symmetric dynamics of the qudit in both  phases (see Fig. \ref{qudit1}  (\ref{qudit2}), representing the dynamics of the the distinguishability in the broken (unbroken) phase, in which the green solid and orange dashed curves are plotted for $\bigg( \theta=\pi/4, \varphi=\pi/2\bigg) $ and $ \bigg(\theta=1.1~(\theta=1.5), \varphi=2.1~(\varphi=0.1) \bigg)$, respectively).

 \par
 Recently, this interesting four-dimensional system,   has been  implemented experimentally  with single photons and a cascaded interferometric setup \cite{bian2020quantum}. In that work,   $ 4\times 4 $ nonunitary evolution operations were realized by six
 beam displacers and  another one is used for state preparation. Moreover, two different measurements, i.e., the projective measurement and the quantum
 state tomography of a four-level system, are carried out at
 the output. 
 We find that our theoretical predictions  are completely    consistent with the experimental results discussed in  \cite{bian2020quantum} in which explicitly reported  that the EP4  ($ \gamma = J $) plays the main role in determining the critical behavior of the qudit.  Moreover, our theoretical analysis shows that the HSS oscillates with period $ T=2\pi/(\sqrt{J^{2}-\gamma^{2}}) $ in the unbroken phase.  This is exactly the period measured experimentally in   \cite{bian2020quantum} for oscillations of the dynamics of the quantum information (as quantified by entropy).  Because the TD is not contractive in this model, these oscillations cannot be interpreted as evidence
 of information backflow from the environment or the signature
 of non-Markovianity. They can be attributed to the non-Hermitian evolution of the system making the initial  hidden information available, as discussed in Sec. \ref{QFIHSS}.

 \section{Conclusions}\label{conclusion}

 We have proposed a powerful and easily computable  witness, based on the Hilbert-Schmidt speed (HSS), which is a special case of quantum statistical speed, to detect  the quantum criticality in systems governed by non-Hermitian (anti-)$\mathcal{PT}\!-$symmetric Hamiltonians.  Surprisingly, our theoretical predictions  can exactly predict the experimental results.
 
 In addition to its conceptual interest, we remark that the HSS-based witness of criticality does not require the diagonalization of the reduced  density matrix of the system.  Hence, as discussed in the paper for a four-dimensional qudit, it can be introduced as a faithful witness for characterizing criticality in high-dimensional (anti-)$\mathcal{PT}\!-$symmetric systems in which computation of  other measures leads to serious challenges. The role of the HSS-based measure in detecting the exceptional points (EPs) at which breaking of (anti-)$\mathcal{PT}\!-$symmetry occurs   has been also analyzed. We have especially illustrated that the   system critical behavior appearing at EPs is similar to that of   (anti-)$\mathcal{PT}\!-$symmetric system  in its broken (unbroken) phase. We stress that our theoretical findings all are  in complete agreement with experimental observations. These results thus indicate that the HSS-based witness can be adopted to exactly identify the parameter values where phase transitions in the physical behavior of non-Hermitian systems occur.

 As an interesting outlook, the introduced HSS-based witness can be employed to characterize controlled speedup of quantum processes in non-Hermitian systems. Such a line of investigation is suggested by recent advances in the context of shortcuts to adiabaticity (STA), which enable us to control a quantum system evolution with no need of slow driving \cite{guery2019shortcuts,funo2020shortcuts,chen2020shortcuts,alipour2020shortcuts}. Strategies for STA are typically engineered by means of non-Hermitian control Hamiltonians \cite{alipour2020shortcuts,Chen2018,Impens2019}. 
 On the basis of this argument and seeing the results presented here, one may thus expect that the HSS measure plays a role in optimizing  STA via non-Hermitian Hamiltonians. This study will be carried on elsewhere.


 Another important feature which should be addressed is the non-contractivity of both trace distance (TD) and HSS   in (anti-)$ \mathcal{PT} \!-$symmetric non-Hermitian systems. 
 Because of this non-contractive behavior, the change in the TD of two arbitrary states can no longer be interpreted as a flow of information between the system and the environment. Therefore, contrary to what happens in Hermitian system \cite{breuer2009measure,jahromi2020witnessing}, both  TD and  HSS   can no longer be used as  general measures or definitions of non-Markovianity in (anti-)$ \mathcal{PT} \!-$symmetric systems. Our results show that the non-contractivity of the Hilbert-Schmidt speed  in a given model may be a signature of the trace distance non-contractivity and its failure in detecting non-Markovianity.
 \par
 We explain that the definition of non-Markovianity remains as an important open question in the theory of non-Hermitian quantum systems. In fact, the distinguishability and HSS oscillations in
 (anti-)$ \mathcal{PT} \!-$symmetric dynamics may be  attributed to  the
 result of the dynamical overlap between the skew eigenstates in the  (anti-)$ \mathcal{PT} \!-$symmetric,  an interesting
 characteristic absent in conventional open quantum systems. In the unbroken (broken) phase, this overlap is nontrivial and leads to
 a beat from the two skew dynamical eigenstates,
 such that the beat period equals the oscillation period. However, in the
 broken (unbroken) phase, the aforementioned overlap becomes trivial, and in addition,
 the amplitudes of the eigenstates monotonically shrink or grow    under quantum evolution, leading to a monotonic decrease in  the 
 distinguishability or the HSS dynamics (see \cite{wang2020experimental} for more details).

 In Figs.~\ref{HssQfi} and \ref{HssQfianti}, we have compared the behaviors of HSS and quantum Fisher information (QFI) associated with the initial phase $ \varphi $ for one-qubit (anti-)$ \mathcal{PT} \!-$symmetric systems. This investigation shows that the HSS and QFI exhibit the same qualitative dynamics in one-qubit systems.  According to the discussion presented in the previous paragraph, we find that by inspecting the  QFI dynamics one cannot exactly detect the non-Markovian evolution of the (anti-)PT-symmetric systems. However, as known, the QFI is  a faithful witness of non-Markovianity in Hermitian systems \cite{lu2010quantum,fujiwara2001quantum}. Therefore, we conclude that the witnesses proposed for Hermitian systems should be reexamined to check their efficiency in non-Hermitian systems.

 Our work thus motivates deeper analyses to clarify the applicability of other witnesses of the non-Markovianity  to detect the memory effects in Hermitian and non-Hermitian systems. Our results also pave the way to further studies on HSS applications in detecting the criticality in high-dimensional non-Hermitian multi-qudit systems.

 \section*{Acknowledgements}
 H.R.J. wishes to acknowledge the financial support of the MSRT of Iran and Jahrom University. 
 The authors would like to thank Franco Nori and Chia-Yi Ju for fruitful discussions and feedbacks.

 \bibliography{Ref}
 
\end{document}